\documentclass[12pt]{article}
\usepackage{cite}
\usepackage[a4paper,hscale=0.82,vscale=0.75]{geometry}
\usepackage[parfill]{parskip}
\usepackage{graphicx}
\usepackage{amssymb}

\usepackage{amsmath}
\usepackage{amsthm}
\usepackage{amscd}
\usepackage[all,cmtip]{xy} 
\numberwithin{equation}{section}
\usepackage{empheq}
\usepackage[british]{babel}
\usepackage[small,nohug]{diagrams} 

\title{A modern view of the classical\\
Herglotz-Noether theorem\vspace{.8cm}}
\author{Ziyang Hu\footnote{\texttt{z.hu@damtp.cam.ac.uk}}\\
D.A.M.T.P.\\
 University of Cambridge}

\newcommand{\pd}{\partial}
\newcommand{\rs}{\mathbb{R}}

\DeclareMathOperator{\diag}{diag}

\DeclareMathOperator{\id}{id}

\newtheoremstyle{shape0}
  {9pt}
  {9pt}
  {}
  {}
  {\bfseries}
  {.}
  {.5em}
  {}

\newtheoremstyle{shape1}
  {9pt}
  {9pt}
  {\it}
  {}
  {\bfseries}
  {.}
  {.5em}
  {}

\newtheoremstyle{shape2}
  {9pt}
  {9pt}
  {}
  {}
  {\itshape}
  {.}
  {.5em}
  {}

\theoremstyle{shape1}
\newtheorem{thm}{Theorem}
\newtheorem{prop}[thm]{Proposition}
\newtheorem{cor}[thm]{Corollary}

\theoremstyle{shape0}

\theoremstyle{shape2}
\theoremstyle{definition}
\newtheorem*{dfn}{Definition}

\begin{document}
\maketitle
\abstract{In this paper we give a new proof, valid for all dimensions, of the classical Herglotz-Noether theorem that all rotational shear-free and expansion-free flows (rotational Born-rigid flows) in Minkowski spacetime are generated by Killing vector fields (isometric flows). This is aimed as an illustration of a general framework for working with problems that can be described as a structure-preserving submersion, which we will describe in a subsequent paper.}
\newpage
\tableofcontents
\newpage

\section{Introduction}
\label{intro}
\subsection{History and previous results}
\label{history}
Man has studied rigid motion since antiquity. The concept of \emph{rigid motion} relevant to us, though, begins only within the framework of Newtonian mechanics. Using modern language, the way of defining such a motion is by considering a body $B$ as a Riemannian manifold (possibly with boundary), the ambient space $M$ also as a Riemannian manifold (usually just Euclidean space) and a map $i$ depending on a parameter $t$ ``time'' taken values in $\rs$, for which the map 
\begin{equation}
  \label{oldf}
  i(t):B\rightarrow M
\end{equation}
obtained by fixing the first argument in $i:\rs\times B\rightarrow M$ is  an isometric embedding for all $t$ in a relevant domain. However, when one comes to relativity, one finds that such a definition no longer works. In particular, in relativity, when $\dim B=\dim M-1$ where $M$ now denotes spacetime, a rigid motion in the Newtonian sense would provide an \emph{absolute} splitting between space and time which is effected, for example, by a moving observer along the flow worldline. 

 Born \cite{born1909}, Herglotz \cite{herglotz1910} and Noether \cite{noether1910} have   formulated a new concept of rigid motion in special and general relativistic spacetime. The definition reads:
\begin{dfn}[Born rigidity]
A body is called rigid if the distance between \emph{every} neighbouring pair of particles, measured orthogonal to the worldlines of \emph{either} of them, remains constant along the worldline.\end{dfn}In this paper, unless otherwise noted, rigidity will always mean Born-rigidity.
Intuitively, this says that \emph{every} element of the body appears rigid in the Newtonian sense to \emph{every} inertial observer instantaneously at rest relative to the element.  Then it comes as a \emph{surprise} when Herglotz \cite{herglotz1910} and Noether \cite{noether1910} independently proved that, in Minkowski spacetime, i.e., special relativity, all such rigid motions which are rotating follows the integral curves of an isometry. The surprise is that in Newtonian spacetime, we are free to move rigid bodies all we want, i.e., the motion has $6$ free parameters, which corresponds exactly to the flowline of a single particle in the rigid body and the orientation of the whole body with respect to this particle, but in the rotational case in relativity, if a rigid body is known at one instant it is determined for all time. The aim of the present paper is, in addition to extending Herglotz and Noether's result, come to terms with such surprises so that they appear more ``natural''.

The first step in achieving this goal is to formulate the concept of rigid motion in Newtonian spacetime on a more equal footing with Born's definition of rigidity in relativity. From a structural point of view, Galilean (Newtonian) \emph{spacetime} is not Euclidean space. It has a canonical coordinate system given by $(t,x,y,z)=(t,\mathbf{x})$ with the allowed transformations of the space are given by the Galilean transformations acting as a matrix
\begin{equation}
\label{lie1}
\begin{pmatrix}
  1\\
  t'\\
  \mathbf{x}'
\end{pmatrix}
=
\begin{pmatrix}
 1&0&\mathbf{0}^{T}\\
 t_{0}&1&\mathbf{0}^{T}\\
 \mathbf{x}_{0}&\mathbf{v}&R
\end{pmatrix}
\begin{pmatrix}
  1\\
  t\\
  \mathbf{x}
\end{pmatrix}
\end{equation}
where $RR^{T}=I$ is an element of the rotation group $SO(3)$\footnote{Since all our considerations are local, in which case only group elements infinitesimally close to the identity are relevant, we can safely ignore all discrete symmetries, and  consider only $SO(3)$, not the whole of $O(3)$. Similar comment applies to all subsequent discussions.}, making the whole group the structurally rather complicated $(SO(3)\ltimes \rs^{3})\ltimes \rs^{4}$. The quantity $t_{0}$ is the translation of time, $\mathbf{x}_{0}$ is the translation of space, $\mathbf{v}$ is the relative velocity of the Galilean transformation, and $R$ is of course the rotation of the coordinate axis. Now, it is rather awkward to talk about metric in Galilean spacetime.  However, if we insist so, then a Galilean transformation preserves, instead of one, two degenerate metrics, which in the canonical coordinates have the explicit expressions
\begin{equation}\label{oddmetric}
g^{(s)}_{ij}=\diag(0,1,1,1),\qquad g^{(t)}_{ij}=\diag(1,\infty,\infty,\infty).
\end{equation}
Obviously, this is not very rigourous, and a better way is to say that, instead of $g^{(t)}_{ij}$, the \emph{contravariant metric} $g_{(t)}^{ij}=(1,0,0,0)$ is the one that really matters. One can then turn the argument around and say that it is these two degenerate metrics $g^{(s)}_{ij}$ and $g_{(t)}^{ij}$ (these have no inverses) that give the space $\rs^{3+1}$ an extra geometrical structure, which makes it Galilean space. The Galilean group can then be \emph{derived} from this definition. We can also see from this that  the intuitively familiar Newtonian space and time is really mathematically speaking much more intricate than Minkowski spacetime.

 It is then easy to see that the Born definition of rigidity in Galilean {spacetime}  coincides with the  definition \eqref{oldf}. Let us then compare it with rigid motion in relativity. Now the isometries are
\begin{equation}
\label{lie2}
\begin{pmatrix}
  1\\
  \mathsf{x}'
\end{pmatrix}
=
\begin{pmatrix}
 1&\mathsf{0}^{T}\\
  \mathsf{x}_{0}&L
\end{pmatrix}
\begin{pmatrix}
  1\\
  \mathsf{x}
\end{pmatrix}
\end{equation}
an element of the Poincar\'e{} group $SO(3,1)\ltimes \rs^{4}$ where now $\mathsf{x}=(ct,\mathbf{x})^{T}$, and $L\eta L^{T}=\eta$ where
\[
\eta_{ij}=\diag(-1,1,1,1).
\]Under the In\"on\"u-Wigner contraction \cite{Inonu:1953p7282} effected by $1/c\rightarrow 0$ where $c$ is the speed of light, we obtain the Galilean group and the degenerate metrics $g_{ij}^{(s)}$ and $g^{ij}_{(t)}$. Why do we get two metrics from one? A fancy statement could perhaps make this clear: in the relativity case, space and time are comparable, but in the Galilean limit, ``time is infinitely more powerful than space'', hence space cannot meddle with the affairs of time, i.e., the allowed transformation for time is a linear function of itself only, with no scaling and no coefficients tied to space, which gives the absolute splitting of space and time. 

Let us now  use orthonormal coframes for Galilean and Minkowski spacetime to write equations for what we have said. Let $\omega^{0}$ denote a time-like unit 1-form and $\omega^{i}$, $i=1,2,3$ denote the spacelike 1-forms such the system of coframe $(\omega^{0},\omega^{i})$ is orthonormal. Notice that in the Minkowski setting, this is only one set of conditions, but in the Galilean case this actually reduces to two sets since we have two metrics. In both cases, to say that a vector field $\mathbf{V}$ describes a rigid motion means that
\begin{equation}\label{rigidcondition}
\mathcal{L}_{\mathbf{V}}\left(\sum_{i=1}^{3}\omega^{i}\otimes\omega^{i}\right)=0
\end{equation}
where $\mathcal{L}_{\mathbf{V}}$ indicates Lie derivative with respect to the vector field $\mathbf{V}$. The integral curves of $\mathbf{V}$ are then the flowlines of this rigid motion, and in relativity they can be interpreted as the worldlines of particles constituent of the flow, with the restriction that $\mathbf{V}$ must be time-like everywhere.
On the other hand, an isometry vector field, i.e., a Killing vector field in Minkowski spacetime must satisfy
\begin{equation}\label{mincondition}
\mathcal{L}_{\mathbf{V}}\left(\sum_{i=1}^{3}\omega^{i}\otimes\omega^{i}-\omega^{0}\otimes\omega^{0}\right)=0.
\end{equation}
whereas in Galilean spacetime we have two conditions
\begin{equation}\label{galcondition}
\mathcal{L}_{\mathbf{V}}\left(\sum_{i=1}^{3}\omega^{i}\otimes\omega^{i}\right)=0,\qquad\mathcal{L}_{\mathbf{V}}(\mathbf{I}^{0}\otimes\mathbf{I}^{0})=0.
\end{equation}
We have denoted the vector field dual to $\omega^{0}$ by $\mathbf{I}^{0}$. Now it is obvious that \eqref{galcondition} implies \eqref{rigidcondition}, but since there is one less equation in \eqref{rigidcondition} than in \eqref{galcondition}, and the other equation in both are identical, that \eqref{rigidcondition} is not sufficient for \eqref{galcondition}. To put in words,
\begin{prop}
  Every isometry in Galilean spacetime determines a rigid motion, but there are rigid motions that fail to be isometries.
\end{prop}
Note that, in the literature sometimes there is a confusion between isometries in Galilean spacetime and isometry in Euclidean space. Considering Euclidean isometry amounts to projecting the vector field down to $\rs^{3}$ using the obvious projection. In this sense, the proposition holds as stated as well, as can be easily shown.

On the other hand, the case in Minkowski spacetime is much more murky. For a start, note that \eqref{mincondition} and \eqref{rigidcondition} contain the \emph{same} number of equations. It is easily shown and widely agreed that \eqref{mincondition} implies \eqref{rigidcondition}, i.e., isometry implies rigidity \cite{pirani1962,pirani1964,Wahlquist:1966p2548,Giulini:2006p66}. Herglotz \cite{herglotz1910} and Noether  \cite{noether1910} independently showed in 1910 that
\begin{thm}[Herglotz, Noether]
  In $(3+1)$-dimensional Minkowski spacetime, every rotational rigid motion is isometric.
\end{thm}

Pirani and Williams \cite{pirani1962} in 1962 gives the theorem firmer foundation by discussing the integrability conditions for such a rigid flow, which also makes possible a new, better proof for the theorem using tensor calculus (the ``debauch of indices''). Indeed, it is the proof by Pirani and Williams, instead of the original proofs by Herglotz and Noether, that is subsequently followed on in the literature. The Pirani-Williams proof was reproduced by Pirani more succinctly in the lectures in \cite{pirani1964} which, in addition,  gives a counterexample showing that in the general Riemannian spacetime of general relativity (4-dimensional), the theorem is false. All these are summarised and reviewed by Giullini in the 2006 review \cite{Giulini:2006p66}, using more modern notation, with emphasis on the importance of this theorem to understanding the foundations of special relativity. All the above investigations are restricted to $(3+1)$ dimensions and are only concerned with the flat case, apart from the counterexample given in the general case. Part of the reason for this lack of extension is the daunting tensor algebra and the obscurity of the geometrical picture.

A different line of development by Wahlquist and Estabrook from 1964 to 1967 \cite{Estabrook:1964p2608,Wahlquist:1966p2548,Wahlquist:1967p2556} used dyadic analysis which, by the choice of method, restricts one to $(3+1)$ dimensions. In this series of papers, it was shown successively the truth of the theorem in Minkowski spacetime, de Sitter and anti-de Sitter spacetimes, and finally, all conformally flat spacetime. However, due to the hermetic language of dyadic analysis  in curved spacetime, the paper was not followed subsequently besides a quote of the statement of the theorem in \cite{tensor}.

\subsection{Relation to the AdS/CFT correspondence}

Recent work by S Bhattacharyya, S Lahiri, R Loganayagam and S Minwalla \cite{Bhattacharyya:2007p31} has used the AdS/CFT correspondence to argue that large rotating black holes in global AdS$_{D}$ spaces are dual to a stationary solution of the relativistic Navier-Stokes equations on $S^{D-2}$, i.e, the conformal boundary, which is the Einstein static universes $S^{D-2}\times \rs$. The fluid is shown to be in dissipationless (shear and expansion free) motion, which (see Section \ref{interpretation} in the present paper) coincides with the condition for Born rigidity. It was shown in \cite{Bhattacharyya:2007p31} that the thermodynamics and the local stress tensor of the solutions obtained are in precise agreement with the thermodynamics and boundary stress tensor of the spinning black holes which, from one perspective, gives strong indication that their methods in particular and the AdS/CFT correspondence in general is a valid theory (though not well understood) that needs to be further clarified. An implicit assumption  they have made in their derivation is that rotational rigid motions in conformally flat spacetime are necessarily locally isometric, occurring in Section 3 of \cite{Bhattacharyya:2007p31}, which is precisely the subject of the present paper.  In simple words: using the AdS/CFT correspondence, rigid motion would tell us something about uniqueness of black holes in the dual description, and a key ingredient to this line of reasoning is understanding the precise relationship between Born-rigidity and isometry. Indeed, the above-mentioned results from AdS/CFT correspondence gives strong indication that the rigidity-isometry equivalence is true in more general settings than the original $(3+1)$ Minkowski spacetime, which we will prove rigourously in this paper.

\subsection{New results and development}
\label{sec:new-results-devel}

In this paper, we shall present a proof that, in Riemannian and pseudo-Riemannian spaces that are homogeneous, rotational rigidity implies isometry. This proof is valid regardless of dimension, and hence extends the classical Herglotz-Noether theorem which is for 4-dimensional Minkowski spacetime.

One aim of this paper is to provide a more down-to-earth illustration of a method which we shall present in a subsequent paper \cite{me:001} for dealing with the general problem of a structure-preserving submersion. In \cite{me:001} we shall also pursue further the problem that we have investigated in this paper: namely, we will prove that the Herglotz-Noether theorem is valid for all conformally flat spacetime, and moreover, remains valid when considered within the framework of conformal geometry.

\section{Riemannian geometry}
\label{math}
Here we will only develop the Cartan theory of generalised spaces necessary for understanding Riemannian geometry. In \cite{me:001} we will give a more general treatment.
\subsection{Coframes}
\label{mf}
Consider the principal bundle of orthonormal frames $P$ over the manifold $M$. The base is of Lorentzian signature, and we will use the Greek indices to denote both spacelike and (unspecified) timelike quantities. When a timelike quantity is specified, the index will always have the value $0$. Latin indices will be reserved for spacelike quantities.

On this bundle, the generalised left invariant Maurer-Cartan form (i.e., Cartan connections) $\omega^{\mu}$ and $\omega^{\mu}{}_{\nu}$ provides a global coframing of $P$. The forms $\omega^{\mu}$ are \emph{semi-basic}, i.e., contraction with any vector along the fibre gives zero: $i_{\mathbf{V}}\omega^{\mu}=0$ if $\mathbf{V}$ projects to $0$ under the canonical projection, while the forms $\omega^{\mu}{}_{\nu}$ are vertical: projection always gives zero. More concretely, if we have a local trivialisation for the bundle as $\rs^{n}\times H$ where $H$ is the isotropy group of $M$, i.e., it is $SO(1,n-1)$, and let $x^{\mu}$ be coordinates of $\rs^{n}$ and let $u^{A}$ coordinatise $H$. Then,
\[
\omega^{\mu}=F^{\mu}{}_{\nu}(x,u)dx^{\nu},\qquad\omega^{\mu}{}_{\nu}=G^{\mu}{}_{\nu A}(x,u)du^{A}.
\]
The forms $\omega^{\mu}{}_{\nu}$ has symmetries in accordance with the Lie algebra of the group in question. In our case, the symmetry is
\[
\omega^{i}{}_{j}=-\omega^{j}{}_{i},\qquad\omega^{i}{}_{0}=\omega^{0}{}_{i}.
\]
Note that we are using what is called the \emph{tangent space indices} for the forms $\omega^{\mu}$ and $\omega^{\mu}{}_{\nu}$ (and later $\mathbf{I}_{\mu}$), in which the indices label forms/vectors, not components of a form or a vector. Hence, $\omega^{\mu}$, which is a form, has its indices upstairs instead of downstairs.

We can raise or lower an index with the fibre metric $\eta$. In short, when a Latin index is raised or lowered, the sign is unchanged, whereas if $0$ is raised or lowered, we need to have a minus sign. We can combine all $\omega^{\mu}$ and $\omega^{\mu}{}_{\nu}$ into a single $\omega$, the left-invariant Maurer-Cartan form with values in the Lie algebra of the group of isometries. Then the structural equation for the group (the flat case) is written
\[
d\omega+\tfrac{1}{2}[\omega,\omega]^{\wedge}=0,
\]
in which the symbol $[\ ,\ ]^{\wedge}$ denotes the composite map
\[
[a\otimes\mathsf{f},b\otimes\mathsf{g}]^{\wedge}=(a\wedge b)\otimes[\mathsf{f},\mathsf{g}]
\]
where $a,b$ are (scalar-valued) differential 1-forms, and $\mathsf{f},\mathsf{g}$ are elements in some Lie algebra $\mathfrak{g}$ with Lie bracket $[\ ,\ ]$.

If a section $s:M\rightarrow P$ is taken, then $s^{*}(\omega^{\mu})$, being linearly related to $dx^{\mu}$, furnish a co-frame for $M$.  Usually we will omit  writing out explicitly the pullback acting on the Maurer-Cartan forms. Hence, \emph{when and only when a section is taken}, all the forms $\omega^{\mu}{}_{\nu}$ can be expressly linearly in terms of the $\omega^{\mu}$.

The first Maurer-Cartan equation is always satisfied (more precisely, it is used as one of the conditions for the definition of the $\omega^{\mu}{}_{\nu}$ in a torsionless geometry):
\begin{equation}
  \label{eq:1}
  d\omega^{\mu}=\omega^{\nu}\wedge\omega^{\mu}{}_{\nu}.
\end{equation}
If the space is Minkowskian, i.e., flat, then the second Maurer-Cartan equation is also satisfied
\begin{equation}
  \label{eq:2}
  d\omega^{\mu}{}_{\nu}=\omega^{\lambda}{}_{\nu}\wedge\omega^{\mu}{}_{\lambda}.
\end{equation}
We should think of the Maurer-Cartan equations as infinitesimal forms of the group composition laws of the group of isometries for a geometry. Indeed, a manifold on which Lie algebra-valued 1-forms can be defined which satisfies the Maurer-Cartan equation is locally  a Lie group with the said Lie algebra.

For general spacetime, we have
\begin{equation}
  \label{eq:3}
  d\omega^{\mu}{}_{\nu}=\omega^{\lambda}{}_{\nu}\wedge\omega^{\mu}{}_{\lambda}+\Omega^{\mu}{}_{\nu}
\end{equation}
where $\Omega^{\mu}{}_{\nu}$ is the curvature 2-form. It measures the extent of the failure of the group of isometries of flat spacetime to act on the spacetime in question as an isometry. It depends only on $\omega^{\mu}$, i.e., a decomposition
\[
\Omega^{\mu}{}_{\nu}=\tfrac{1}{2}R^{\mu}{}_{\nu\rho\lambda}\,\omega^{\rho}\wedge\omega^{\lambda}
\]is unique \textit{without using a section}. If a section is taken, then $R^{\mu}{}_{\nu\rho\lambda}$ becomes the components of the usual Riemann curvature tensor in the frame chosen. The condition $\Omega^{\mu}{}_{\nu}=0$ is used as an integrability condition: if it is satisfied, the space is locally Minkowskian.

We have discussed curvature as a measure of the failure of the group of isometries of flat spacetime (i.e., $ISO(n-1,1)$) to act as a group of isometries for the spacetime in question. Hence the centre stage is taken by the group. However, a sphere $S^{n}$, for example, has a perfectly valid effective and transitive action given by $SO(n)$, with isotropy group given by $SO(n-1)$. Our previous approach only requires cosmetic changes if we were to use the Maurer-Cartan equations for $SO(n)$ to define flatness (in which case Euclidean spaces are curved). We can formalise this as follows: for a \emph{mutation} of model, we mean the change in the definition of $\Omega'^{\mu}{}_{\nu}$ by
\begin{equation}
\label{mutation}
d\omega^{\mu}{}_{\nu}=\omega^{\lambda}{}_{\nu}\wedge\omega^{\mu}{}_{\lambda}-\kappa\omega^{\mu}\wedge\omega_{\nu}+\Omega'^{\mu}{}_{\nu}
\end{equation}
where the constant $\kappa$ is a parameter for the model. For $\kappa>0$, the model is based on the de Sitter spacetime, and $\kappa<0$, the model is based on the anti-de Sitter spacetime. Note that the forms $\omega^{\mu}$ scales proportionally to the length of a vector, whereas $\omega^{\mu}{}_{\nu}$ does not, hence by appropriate scaling, we only need to consider the cases of $K=\{0,-1,1\}$. $\Omega'^{\mu}{}_{\nu}$ is still an integrability condition, i.e., if it vanishes, then it is locally identical to the chosen homogeneous space. All these  are connected by the fact that any homogeneous space can be represented the quotient of a Lie group by a subgroup, and the metric tensor is the remnant of the Cartan-Killing metric on the Lie group for the remaining Lie algebra generators not discarded (usually non-compact generators if the space is non-compact).

We will use bold letters to denote vectors. The letter $\mathbf{M}$ is exceptional, however, as we will use it to denote a variable point in the manifold (this is sometimes confusingly called ``thinking of a space as an affine space''). The precise meaning is this: in flat spacetime, once an origin is chosen, we can represent every point as a vector. Now general spacetimes always has an osculating flat spacetime, and hence, \emph{infinitesimally}, such a representation is still valid. See \cite{Cartan:1983p6178,Cartan:2001p6159} for more details.  The symbols $\mathbf{I}_{\mu}$ will denote an orthonormal frame. Choosing the dual frame to be $\omega^{\mu}$ is equivalent to taking a section, in the principal bundle, and we have the following identities
\begin{equation}
  \label{eq:4}
  d\mathbf{M}=\omega^{\mu}\otimes\mathbf{I}_{\mu},\qquad d\mathbf{I}_{\mu}=\omega^{\nu}{}_{\mu}\otimes\mathbf{I}_{\nu}.
\end{equation}
We will be using moving (co-)frames most adapted to the problem at hand. This boils down to specifying parts, or all, of the frames $\mathbf{I}_{\mu}$ and $\omega^{\mu}$. The 0th order moving frame is just the principal bundle itself, with all basis vectors for the moving frame arbitrary, and the 1st order moving frame is one in which we impose certain relations on the bundle by means of one condition. This is, in effect, a reduction of the principal bundle to one with a smaller fibre with a smaller group acting on the right. Unless all basis vectors are specified, we have a non-trivial action on the non-trivial fibre of the principal bundle. It is useful to know how a change of section in the bundle affects various quantities. A change of section is, viewed in the pedestrian way, a non-constant rotation (we are using the word in a generalised sense in the Lorentzian setting to include boosts) $A^{\mu}{}_{\nu}$ acting on the basis vectors by
\[
\mathbf{I}_{\mu}'=\mathbf{I}_{\nu}A^{\nu}{}_{\mu},
\]
hence
\begin{equation}
  \label{eq:11}
  \omega'^{\mu}=(A^{-1})^{\mu}{}_{\nu}\omega^{\nu}
\end{equation}
and we can calculate the important result
\begin{equation}
  \label{eq:5}
\omega'^{\mu}{}_{\nu}=(A^{-1})^{\mu}{}_{\lambda}\omega^{\lambda}{}_{\rho} A^{\rho}{}_{\nu}+(A^{-1})^{\mu}{}_{\lambda}dA^{\lambda}{}_{\nu}.
\end{equation}
We remark that the forms $\omega^{\mu}{}_{\nu}$, when all restrictions on the moving frame is imposed, furnish a complete set of linearly independent differential form for the problem at hand, and hence a complete set of functionally independent differential invariants can be derived from them.

\subsection{Covariant derivatives}
\label{connection}
On the principal bundle, we have the forms $\omega^{\mu}$ and $\omega^{\mu}{}_{\nu}$ as basis for the cotangent space, whereas $\mathbf{I}_{\mu}$ and $\mathbf{I}_{\mu}{}^{\nu}$, the dual quantities, are basis for the tangent space. All other quantities, \emph{having indices or not}, are viewed as \emph{scalar functions} on the principal bundle which changes according to certain representation of fibre (a Lie group) under right action by that Lie group. If a product of such symbols is such that there are no non-contracted indices, then once pulled back to the base by a section, it becomes a scalar. Otherwise, the free indices becomes tensor indices in the usual fashion. In trying to solve our problem, we obtained equations relating to their derivatives. The question is: how to differentiate such scalars \emph{on the bundle} with respect to directions \emph{on the base} in a ``covariant'' fashion. We now briefly reviews the relevant formalism. The notation looks similar to that in tensor calculus, but it is important to bear in mind that on the bundle, they are only scalar functions.

Consider a bundle $\pi:E\rightarrow M$ which is a rank $r$ vector bundle over $M$. Let $s:M\rightarrow E$ be a section. The differential of $s$ is a mapping of tangent space $ds_{x}:T_{x}M\rightarrow T_{s(x)}E$. Another way to say this is that $ds\in C^{\infty}(TM,E)$. Note that the range is too big: the whole of $E$, whereas we actually want the range to be $E_{x}=\pi^{-1}(x)$.

Suppose we have a trivialisation over the open set $U\subset M$ such that $E|_{U}=U\otimes\rs^{r}$. In this trivialisation, we have $s=s^{i}\mathbf{e}_{i}$, as in our trivial example. Then $d^{U}s=ds^{i}\otimes \mathbf{e}_{i}$. For another trivialisation over $V$, we have the basis $\mathbf{\tilde e}_{i}=A^{j}{}_{i}(x)\mathbf{e}_{j}$ as before, but now the transformation $A^{j}{}_{i}$ is varying over $U\cap V$. Then
\begin{equation*}
  d^{V}s=d\tilde s^{i}\otimes \mathbf{\tilde e}_{i}=d(s^{i}(A^{-1})^{j}{}_{i})\otimes\mathbf{\tilde e}_{j}=ds^{i}\otimes\mathbf{e}_{i}-s^{i}(A^{-1})^{j}{}_{k}dA^{k}{}_{i}\otimes\mathbf{ e}_{j}
\end{equation*}
where we have used $dA^{-1}=A^{-1}dA\, A^{-1}$, considered as a matrix equation. If $E$ is a vector bundle associated to a principal bundle with a right $H$ action, i.e., our transformation matrix $A$ lies in a certain matrix Lie group, then of course $A^{-1}dA$ is the left-invariant Maurer-Cartan form for the group $H$. What we want to do is to eliminate this difference $(A^{-1}dA)s$ by tweaking with our definition so that a \emph{covariant derivative} of the section $\nabla s$ is well-defined.

The tangent space $T_{p}E$ has a distinguished \emph{vertical subspace}
\[
\mathcal{V}_{p}E=\{\mathbf{v}\in T_{p}E\mid\pi_{*p}(v)=0\}.
\]
and this subspace is naturally isomorphic to the fibre $E_{\pi(p)}$. If we can unambiguously choose a projection $\Pi_{\text{vert}}$ onto this vertical subspace, then
\[
\nabla s=\Pi_{\text{vert}}\circ d^{U}s
\]
will be well-defined (independent of the choice of trivialisation $U$). From linear algebra, we know that this unique decomposition is possible if and only if we also specify a complementary horizontal subspace, then every vector $\mathbf{v}$ can be decomposed uniquely as $\mathbf{v}=\mathbf{v}_{\text{vert}}+\mathbf{v}_{\text{hor}}$. A choice of such a subspace is a \emph{connection on vector bundle}, and the definition we have given for $\nabla s$ is the \emph{covariant derivative} with respect to this connection for the section $s$.

Specialising to our case of interest, we have a coframing of the principal bundle $P$ of orthonormal frames over the manifold $M$ given by the forms $\omega^{\mu}$ and $\omega^{\mu}{}_{\nu}$. The forms $\omega^{\mu}$ are horizontal and $\omega^{\mu}{}_{\nu}$ are vertical and moreover, $\omega^{\mu}{}_{\nu}$ are the Maurer-Cartan forms in the fibre. Hence they justify the name of being a Cartan \emph{connection}. Let us see how this works in practise by considering the simplest example of a vector-valued function on $M$---a vector field. For a section $s:M\rightarrow P$, $s^{*}(\omega^{\mu})$ furnish an orthonormal coframe on $M$ (as usual, we will omit the pullback sign) and the dual frame is denoted by $\mathbf{I}_{\mu}$, as before. For a vector $\mathbf{V}=V^{\mu}\mathbf{I}_{\mu}$, we have
\begin{equation*}
  \nabla \mathbf{V}=dV^{\mu}\otimes\mathbf{I}_{\mu}+V^{\mu}d\mathbf{I}_{\mu}=(dV^{\mu}+\omega^{\mu}{}_{\nu}V^{\nu})\otimes\mathbf{I}_{\mu}=(\nabla V^{\mu})\otimes \mathbf{I}_{\mu}.
\end{equation*}
In the above expression, $\nabla V^{\mu}$ is an ordinary differential 1-form. We can expand it as
\[
\nabla V^{\mu}=V^{\mu}{}_{;\nu}\omega^{\nu}
\]
where we come to the traditional notation for using semicolon to denote covariant differentiation. Similarly, for a 1-form $\theta=\theta_{\mu}\omega^{\mu}$, we have
\begin{equation*}
  \nabla\theta
=(d\theta_{\mu}-\omega^{\nu}{}_{\mu}\theta_{\nu})\otimes\omega^{\mu}
=(\nabla \theta_{\mu})\otimes \omega^{\mu}
=\theta_{\mu;\nu}\omega^{\nu}\otimes\omega^{\mu}.
\end{equation*}

\section{Flow in an orthonormal frame}
\label{flow}
Let us now start our investigation proper. First, consider  a general flow, without any further restriction. Let $\omega^{\mu}$ be an orthonormal coframing of the manifold, and further suppose that it is first  adapted to the flow such that $\mathbf{I}_{0}$ dual to $\omega^{0}$ is everywhere proportional to the flow vector field. In the principal bundle with fibre $G$, where $G=SO(1,n-1)$ is the isotropy group of the whole manifold, we have, as usual, the Maurer-Cartan relations
\[
d\omega^{0}=\omega^{i}\wedge\omega^{i}{}_{0},\qquad d\omega^{i}=\omega^{j}\wedge\omega^{i}{}_{j}+\omega^{0}\wedge\omega^{i}{}_{0}.
\]However, since $\omega^{0}$ is now completely determined by the flow, not all of $G$ is relevant here. Indeed, the only group elements of $g$ that matters are those that preserves $\omega^{0}$ and $\omega^{i}$ separately, i.e., the fibre has be reduced to $G'=SO(n-1)$. This means that $\omega^{0}{}_{i}$ and $\omega^{i}{}_{0}$ are now semi-basic and can be written as
\begin{equation}
  \label{eq:6}
\omega^{0}{}_{i}=\omega^{i}{}_{0}=K_{i}\omega^{0}+M_{ij}\omega^{j},
  \end{equation}
where $M_{ij}$ and $K_{i}$ are \emph{functions} on the principal bundle, and we have used the symmetry relation of the group of $SO(1,n-1)$. It is perhaps easier to see this by considering explicit group transformations:
 for a transformation given by $A^{\mu}{}_{\nu}$,
\[
\omega'^{0}{}_{i}=(A^{-1})^{0}{}_{\lambda}\omega^{\lambda}{}_{\rho} A^{\rho}{}_{i}+(A^{-1})^{0}{}_{\lambda}dA^{\lambda}{}_{i}.
\]
Due to the restriction that $\omega^{0}$ is completely determined, however,
\[
A^{0}{}_{\lambda}=\delta^{0}{}_{\lambda},\qquad(A^{-1})^{0}{}_{\lambda}=\delta^{0}{}_{\lambda}
\]
so
\begin{equation}
\label{eq:10}
\omega'^{0}{}_{i}=\omega^{0}{}_{j} A^{j}{}_{i}.\end{equation}
Since $A^{j}{}_{i}$ are arbitrary elements of $SO(n-1)$, this transformation law shows that, after reduction, the forms $\omega^{0}{}_{i}$ are no longer independent and are expressible in terms of the semi-basic forms.

Let us also consider the full set of Maurer-Cartan relations. In the bundle of $G$, we have
\begin{align}
  \label{eq:19}
  d\omega^{0}&=\omega^{i}\wedge\omega^{0}{}_{i},\\
  d\omega^{i}&=\omega^{j}\wedge\omega^{i}{}_{j}+\omega^{0}\wedge\omega^{i}{}_{0}.
\end{align}
In the bundle of $G'$, on the other hand, we have
\begin{align}
  \label{eq:20}
  d\omega^{0}&=K_{i}\omega^{i}\wedge\omega^{0}+M_{ij}\omega^{i}\wedge\omega^{j},\\
\label{eq:20a}  d\omega^{i}&=\omega^{j}\wedge\omega^{i}{}_{j}+M_{ij}\omega^{0}\wedge\omega^{j}.
\end{align}
In the Cartan equivalence method framework, \eqref{eq:20} and \eqref{eq:20a} are the structural equation for the coframe on the base manifold $M$, where $\omega^{i}{}_{j}$ is the Maurer-Cartan form on the \emph{Lie group} $SO(n-1)$. Hence we see that both $M_{ij}$ and $K_{i}$ are torsions of the problem, but the torsion in \eqref{eq:20a} can be absorbed by defining
\[
\widetilde\omega^{i}{}_{j}=\omega^{i}{}_{j}-M^{i}{}_{j}\omega^{0}
\]
after which we have
\begin{align}
\label{esstor}  d\omega^{0}&=K_{i}\omega^{i}\wedge\omega^{0}+M_{ij}\omega^{i}\wedge\omega^{j},\\
  d\omega^{i}&=\omega^{j}\wedge\widetilde\omega^{i}{}_{j}.
\end{align}
We see that, in \eqref{esstor}, both $K_{i}$ and $M_{ij}$ are \emph{essential torsions} that do not depend explicitly on the group parameters at this stage, hence they are the differential invariants of the problem (notice, however, we need to go into the bundle to define them). Later, we shall also discuss how to interpret this absorption of torsion geometrically in the rigid flow case.

Is this the whole story? Do we have two \emph{independent} differential invariants $M_{ij}$ and $K_{i}$, i.e., in terms of scalar-valued functions, $n(n-1)$ of them? This turns out to be not the case. To resolve this, we need to prolong the problem into the bundle, and then we have additional exterior differential equations to consider. On the bundle of $G$, we have, additionally,
\begin{align}
  d\omega^{0}{}_{i}&=\omega^{j}{}_{i}\wedge\omega^{0}{}_{j},\\
  d\omega^{i}{}_{j}&=\omega^{k}{}_{j}\wedge\omega^{i}{}_{k}+\omega^{0}{}_{j}\wedge\omega^{i}{}_{0},\end{align}
and on the bundle of $G'$, we have, \emph{by reduction} of the previous two equations,
\begin{align}
\label{gen1eq}  d\omega^{i}{}_{j}&=\omega^{k}{}_{j}\wedge\omega^{i}{}_{k}+(K_{j}M_{ik}-K_{i}M_{jk})\omega^{0}\wedge\omega^{k}+M_{jk}M_{il}\omega^{k}\wedge\omega^{l},\\
  d(K_{i}\omega^{0}+M_{ij}\omega^{j})&=K_{j}\omega^{i}{}_{i}\wedge\omega^{0}+M_{jk}\omega^{j}{}_{i}\wedge\omega^{k}.
\end{align}
However, only the first one, \eqref{gen1eq}, counts, since this relates how $\omega^{i}{}_{j}$ behaves under exterior differentiation (or rather, it relates how $\widetilde \omega^{i}{}_{j}$ behaves, which is what matters here). The second one is then interpreted as a \emph{constraint} on $M_{ij}$ and $K_{j}$, so these two quantities are not really independent!

We will, in the following discussions, use extensively the following construction made possible by the flow. The flow, being a one dimensional distribution, is trivially integrable and we have a foliation on the space $M$. By considering each leaf as an equivalence class, we obtain a vector bundle $\varpi:M\rightarrow B$, where $B$ is the leaf space or body space. The fibres of this bundle are all isomorphic since they are all one dimensional. \footnote{This is actually an assumption: it is possible to arrange the manifold such that closed timelike curves exist, and then the fibres would be of two kinds: lines and circles. However, closed timelike curves are certainly unphysical. Furthermore, since everything we do will be local in character, this assumption does not place a severe restriction on our results.}

\subsection{Rigid flow}
Using Cartan's classical way of writing \cite{Cartan:1983p6178,Cartan:2001p6159},  let $\mathbf{M}$ denote a point in $M$, then we have
\[
d\mathbf{M}=\omega^{\mu}\otimes\mathbf{I}_{\mu}=\omega^{0}\otimes\mathbf{I}_{0}+\omega^{i}\otimes\mathbf{I}_{i}.
\]
The line element in $M$ is given by the quadratic form
\begin{equation}
  \label{eq:7}
  ds^{2}_{M}=\sum\omega^{\mu}\otimes\omega^{\mu}.
\end{equation}
Suppose the displacement $\delta$ of $\mathbf{M}$ is orthogonal to the flow. Then
\[
\delta\mathbf{M}=\omega^{i}(\delta)\mathbf{I}_{i},\qquad\omega^{0}(\delta)=0.
\]
The line element for such a displacement is therefore
\[
\delta s^{2}_{B}=\sum\omega^{i}\otimes\omega^{i}.
\]
Let $\mathbf{V}$ be a vector field along this flow, i.e., $\mathbf{V}=\lambda\mathbf{I}_{0}$ for a scalar $\lambda$ which can vary from point to point. Suppose that the flow is a rigid motion, then
\begin{equation}
\label{bornc}
\mathcal{L}_{\mathbf{V}}\delta s^{2}_{B}=0.
\end{equation}
But, since $\omega^{i}(\mathbf{I}_{0})=0$ (orthonormal frame), we have $\omega^{i}(\mathbf{V})=0$, so
\begin{align}
\label{born}
\mathcal{L}_{\mathbf{V}}\omega^{i}&=i_{\mathbf{V}}d\omega^{i}+d(i_{\mathbf{V}}\omega^{i})\\
&=i_{\mathbf{V}}(\omega^{j}\wedge\omega^{i}{}_{j}+\omega^{0}\wedge\omega^{i}{}_{0}).
\end{align}
If a coframing is chosen (i.e., a section is determined), the only independent 1-forms are $\omega^{\mu}$. Therefore we have the following further expansion
\[
\omega^{i}{}_{j}=A^{i}{}_{jk}\omega^{k}+B^{i}{}_{j}\omega^{0}
\]
where both $A^{i}{}_{jk}$ and $B^{i}{}_{j}$ are antisymmetric with respect to the first two indices. Continuing calculating \eqref{born},
\begin{align*}
  \mathcal{L}_{\mathbf{V}}\omega ^{i}&=i_{\mathbf{V}}(\omega^{j}\wedge B^{i}{}_{j}\omega^{0}+\omega^{0}\wedge M^{i}{}_{k}\omega^{k})\\
&=\lambda(M^{i}{}_{j}-B^{i}{}_{j})\omega^{j}.
\end{align*}
Substitute into \eqref{bornc}, we have
\begin{equation}
\label{key}
\mathcal{L}_{\mathbf{V}}\delta s^{2}=\sum_{i,j}\lambda(M_{ij}-B_{ij})(\omega^{i}\otimes\omega^{j}+\omega^{j}\otimes\omega^{i}).
\end{equation}
Since $B_{ij}$ is already antisymmetric, it does not contribute to the sum. Hence, we have the following
\begin{prop}
\label{firstprop}
  The flow is rigid if and only if $M_{(ij)}=0$.
\end{prop}
This is good news---we killed a lot of the degrees of freedom in $M_{ij}$. Notice how $\lambda$ comes in---see our previous discussion on the potential loss of information when passing into an orthonormal frame. 
Note that our derivation does not depend on any particular section of the reduced bundle being taken, and hence it holds in the bundle. Further note that the value of $\lambda$ plays no part in this condition. This will be different when we consider isometric flow.

The above calculation also shows that rigid flow preserves inner product orthogonal to the flow. This condition is sufficiently interesting and applicable to a wide range of problems and it has a special name in geometry: Riemannian submersion. Its significance is given by \eqref{bornc}, which can be stated as
\begin{cor}
  A rigid flow is equivalent to a Riemannian submersion on the space $M$, and endows the leaf space $B$ with a well-defined Riemannian metric given by the line element $\delta s^{2}_{B}$.
\end{cor}
Indeed, one can put this observation into an abstract setting and define \emph{structure-preserving submersions} that preserves a certain subgroup of the isotropy group, and hence generalise Riemannian submersion to all geometries. We shall consider this more general problem in a subsequent paper.

In the following, we will denote operators that are defined on the leaf space by a subscript or superscript with the letter $B$, as we have already done for $\delta s_{B}^{2}$.

\subsection{Conditions for isometric rigid flow}
We will only consider isometric flows that are already rigid.  Hence, instead of calculating the Lie derivative of the whole \eqref{eq:7}, in view of the rigidity of the flow the condition reduces to
\[
\mathcal{L}_{\mathbf{V}}(\omega^{0}\otimes\omega^{0})=0.
\]
Calculating as before,
\begin{align*}
  \label{eq:8}
 \mathcal{L}_{\mathbf{V}}\omega^{0}&=i_{\mathbf{V}}d\omega^{0}+d(i_{\mathbf{V}}\omega^{0})\\
&=i_{\mathbf{V}}(\omega^{i}\wedge\omega^{0}{}_{i})+d\lambda\\
&=i_{\mathbf{V}}(\omega^{i}\wedge K_{i}\omega^{0})+\lambda_{,0}\omega^{0}+\lambda_{,i}\omega^{i}\\
&=\lambda_{,0}\omega^{0}+(\lambda_{,i}-\lambda K_{i})\omega^{i}
\end{align*}
therefore
\begin{equation}
\label{zero}
\mathcal{L}_{\mathbf{V}}(\omega^{0}\otimes\omega^{0})=2\lambda_{,0}\omega^{0}\otimes\omega^{0}+(\lambda_{,i}-\lambda K_{i})(\omega^{i}\otimes\omega^{0}+\omega^{0}\otimes\omega^{i}).
\end{equation}
Since $\omega^{\mu}$ are arbitrary, we must have
\begin{equation}
  \label{eq:9}
  \lambda_{,0}=0,\qquad\lambda_{,i}=\lambda K_{i}.
\end{equation}

Let us think about what we have  here. If a flow is given explicitly, the quantity $K_{i}$, being differential invariant for the problem at hand, can be calculated easily. The quantity $\lambda$, however, is completely arbitrary as long as it satisfies \eqref{eq:9} for the isometry condition to hold. This means that we really need to eliminate $\lambda$ and get a condition that should be satisfied by $K_{i}$. (On the other hand, given $K_{i}$, we can reconstruct the flow by solving \eqref{eq:9} to get $\lambda$, hence in this case, our reduction to an orthonormal frame does not lose us any information---see above discussions about the significance of $\lambda$. This is not generally true, though, in that in more general problems of flows, reduction to an orthonormal frame could lose  information.)

Using an instantaneous frame in which $dx^{i}$ corresponds to $\omega^{i}$ and $dx^{0}$ corresponds to $\omega^{0}$,  we can write the second equation as
\[
K_{i}=\frac{\pd \log\lambda}{\pd x^{i}}
\]
this shows that the co-vector valued differential 1-form is locally exact
\[
K\equiv K_{i}\omega^{i}+0\cdot\omega^{0}=d\log\lambda
\]
for this to hold locally, it is necessary and sufficient that
\[
dK=0,\qquad K_{[i;j]}=0
\]by the Poincar\'e{} lemma, and
\[
K_{i} \text{ remains constant on the flow}
\]
since we need to ensure that $\lambda_{,0}=0$.
Note that the derivative, denoted by a semicolon, can be \emph{any} torsion-free covariant derivative. We write it in this way to anticipate what is to come. Moreover, $K_{i,j}$ cannot be well-defined, though $K_{i;j}$ and $K_{[i,j]}$ are. Hence we have the condition
\begin{prop}
\label{rfif}
  A rigid flow is also an isometric flow if and only if $K_{[i;j]}=0$ and $K_{i}$ remains constant on the flow.
\end{prop}

\subsection{Relations to fluid mechanics}
\label{interpretation}
First, we will investigate how $M$ and $K$ are related to quantities usually defined in fluid mechanics. Assume the moving frame is adapted to the flow. The, the covariant derivative of the vector $\mathbf{I}_{0}$, i.e., the flow line, is
\[
\nabla \mathbf{I}_{0}=(\nabla\delta^{\mu}_{0})\mathbf{I_{\mu}}=\delta^{\nu}{}_{0}\omega^{\mu}{}_{\nu}\otimes \mathbf{I}_{\mu}=K^{i}\omega^{0}\otimes \mathbf{I}_{i}+M^{i}{}_{j}\omega^{j}\otimes \mathbf{I}_{i}.
\]
Pre-multiplying (interior multiplication) with $\mathbf{I}_{0}$ gives $K^{i}\mathbf{I}_{i}$, which is the \emph{acceleration} of the flow. Restricting to the orthogonal subspace, we are left with only $M_{ij}$. As a rank two tensor, it decomposes into three parts (invariant subspaces): the anti-symmetric part, the traceless symmetric part and the trace part. These are called in the fluid literature the \emph{angular velocity}, \emph{shear velocity} and \emph{expansion velocity} of the flow, respectively. Hence a flow is rigid if and only if it is shear and expansion free, and it is in addition isometric if and only if the acceleration is a closed 1-form (when written as a covector, of course).

We remark that there are two more related interpretations of the form $M$ in sections \ref{int} and \ref{torsion}, as the integrability tensor for the base manifold $B$ and the torsion of the Cartan connection $\omega^{i}{}_{j}$ on $B$.

The interpretation of $K$ in terms of classical differential geometry of submanifolds is also possible. Suppose in the space $N$ we have a submanifold $S$, i.e., an embedding $i:S\rightarrow N$ which also defines the Riemannian metric on $S$ by $i^{*}(ds^{2}_{N})$, with $\dim N=n$, $\dim S=s$. The \emph{second fundamental form} is defined on the principal bundle by
\[
\text{II}=\omega^{A}{}_{J}\otimes\omega^{J}\otimes \mathbf{I}_{A}
\]
where indices such as $I, J,\dots$ from the second part of the Roman alphabet denote directions tangent to the submanifold, whereas $A, B,\dots$ from the first part denote directions normal to the submanifold. (In our case, this can be written succinctly as $\text{II}=-d\mathbf{M}\cdot\nabla \mathbf{I}_{0}$.)
Due to integrability of the embedding, this can be expanded as
\[
\text{II}=h^{A}_{IJ}\omega^{I}\otimes\omega^{J}\otimes\mathbf{I}_{A},\qquad h^{A}_{IJ}=h^{A}_{JI}
\]
and the quantity $h^{A}_{IJ}$ descends onto the manifold $N$, i.e., it is constant along fibres of the principal bundle over $N$.
In our case, the submanifold are the 1-dimensional leaves due to the flow, and we have
\[
h^{A}_{IJ}\rightarrow h^{i}_{00}=K^{i}
\]
which describes the extrinsic geometry of \emph{one leaf alone}. The symmetry of the quantity $h^{A}_{IJ}$ is automatically satisfied due to trivial integrability. Note that the definition of $M$ requires the whole flow, i.e., if only a single leaf is given, we can only calculate $K$ but not $M$ (but the condition of isometric flow---Proposition \ref{rfif}, of course depends on how the leaves are stacked together in $M$).

\section{Intrinsic geometry and integrability of rigid flow}
\label{intrinsic}
Up until now we have not utilised the additional metric structure on $B$, which will make clear further constraints on $M_{ij}$ and $K_{i}$. If we denote a certain leaf of the manifold $M$ by $L_{p}$ which is identified with a point $p\in B$, we have the following relations
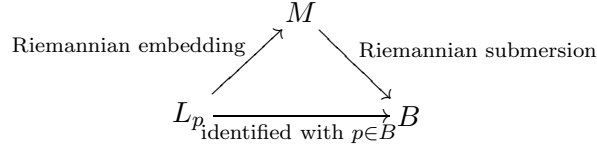
\[
\xymatrix{
&M\ar[dr]^{\text{Riemannian submersion}}&\\
L_{p}\ar[ur]^{\text{Riemannian embedding}}\ar[rr]_{\text{identified with }p\in B}&&B
}
\]
All three spaces $M$, $L_{p}$ and $B$ are endowed with (pseudo-)Riemannian structures, with the ones on $L_{p}$ and $B$ positive definite, and both are derived from the metric on $M$ and the flow. What we will do now is to relate the intrinsic geometry of these three spaces together and derive a series of constraints on these geometries.
\subsection{Gauss and Codazzi equations}
\label{gceq}

The question we will try to answer now is: what restrictions on geometry does the immersion (embedding) of the leaf $L_{p}$ into $M$ give us? It is well known from classical differential geometry of surfaces that the necessary and sufficient condition for an embedding is that the Gauss and Codazzi equations are satisfied.

The leaves is in a distributional sense defined by the system of $(n-1)$ Pfaffian equations
\[
\omega^{i}=0.
\]This distribution is trivially integrable since it is of codimension 1. For completeness, let us still try to derive the Gauss equation, use Cartan's second structural equation
\[ d\omega^{\mu}{}_{\nu}=\omega^{\lambda}{}_{\nu}\wedge\omega^{\mu}{}_{\lambda}+\tfrac{1}{2}R^{\mu}{}_{\nu\rho\lambda}\,\omega^{\rho}\wedge\omega^{\lambda}.
\]
By setting $\mu=\nu=0$, we have
\[
\omega^{\lambda}{}_{0}\wedge\omega^{0}{}_{\lambda}=0
\]
which is identically satisfied. Next, the Codazzi equation. Set $\mu=0,\nu=i$, we have
\[ d\omega^{0}{}_{i}=\omega^{\lambda}{}_{i}\wedge\omega^{0}{}_{\lambda}+\tfrac{1}{2}R^{0}{}_{i\rho\lambda}\,\omega^{\rho}\wedge\omega^{\lambda}
\]
The Codazzi equation is obtained by restricting to $L_{p}$. Of course, since our leaf is one-dimensional, interior multiplication twice by $\mathbf{I}_{0}$ gives zero, so the Codazzi equation is also identically satisfied. Hence, the leaf geometry does not give us any new condition on the geometry.

\subsection{(Non-)integrability of the base space}
\label{int}
Having done with the upward arrow of the diagram at the beginning of this section, let us now come to the downward arrow, which contains the majority of the information we need to solve our problem. Note that the direction of the arrow indicates that the space $B$  \emph{cannot} be considered as a submanifold of $M$ in general. Now we will derive the necessary and sufficient condition for $B$ to be a submanifold of $M$.

Let us choose, on the space $B$, a section of the principal bundle (fibre $SO(3)$) with components $\widetilde\omega^{i},\widetilde\omega^{i}{}_{j}$. On the  reduced bundle with fibre $SO(3)$ on $M$, for a section, we have the forms $\omega^{\mu},\omega^{\mu}{}_{\nu}$ as before, with
\[
\omega^{0}{}_{i}=K_{i}\omega^{0}+M_{ij}\omega^{j}.
\]
Suppose that there is a map $i:B\rightarrow M$ such that $\varpi\circ i=\id_{B}$ (recall that $\varpi$ is the projection map $M\rightarrow B$ adapted to the foliation). Considering the Cartesian product space $M\times B$. If the system of Pfaffian equations
\begin{equation}
  \label{eq:12}
\widetilde\omega^{i}=\omega^{i},\qquad\widetilde\omega^{i}{}_{j}=\omega^{i}{}_{j}
\end{equation}
is integrable in the Frobenius sense, i.e., all differential identities derived from \eqref{eq:12} are consequences of the algebraic identity \eqref{eq:12}, then by Frobenius theorem (formulated in the distribution and differential form language), in the product space there is a integral manifold for the distribution defined by the Pfaffian forms
\[
\theta^{i}=\widetilde\omega^{i}-\omega^{i},\qquad\theta^{i}{}_{j}=\widetilde \omega^{i}{}_{j}-\omega^{i}{}_{j}
\] In other words, we need to check that
\begin{equation}
  \label{eq:13}
  d\widetilde\omega^{i}=d\omega^{i},\qquad d\widetilde\omega^{i}{}_{j}=d\omega^{i}{}_{j}.
\end{equation}
From the first equation of \eqref{eq:13}, we have
\[
\widetilde\omega^{i}{}_{k}\wedge\widetilde\omega^{k}=\omega^{i}{}_{k}\wedge\omega^{k}+\omega^{i}{}_{0}\wedge\omega^{0}.
\]
Using \eqref{eq:12} algebraically, we have
\[
\omega^{i}{}_{0}\wedge\omega^{0}=0
\]
On substitution from \eqref{eq:6}, we have
\[
M_{ij}\omega^{j}\wedge\omega^{0}=0
\]
hence, a necessary condition for integrability is that\[M_{ij}\omega^{i}\wedge\omega^{j}=0.\]
\begin{cor}
\label{fcor}
  If a rigid flow is undergoing rotational motion, then it is never possible to describe the flow as a motion of a certain spacelike hypersurface with definite Riemannian structure in the total spacetime.
\end{cor}
The condition $M_{ij}=0$ is not in itself sufficient: we also need the second equation of \eqref{eq:13}. We have
\[
\widetilde\omega^{k}{}_{j}\wedge\widetilde\omega^{i}{}_{k}+\widetilde\Omega^{i}{}_{j}=
\omega^{k}{}_{j}\wedge\omega^{i}{}_{k}+\omega^{0}_{j}\wedge\omega^{i}{}_{0}+\Omega^{i}{}_{j}
\]
or
\[
\widetilde\Omega^{i}{}_{j}=
\omega^{0}{}_{j}\wedge\omega^{i}{}_{0}+\Omega^{i}{}_{j}.
\]Now $\omega^{0}{}_{i}=K_{i}\omega^{0}$, so $\widetilde\Omega^{i}{}_{j}=\Omega^{i}{}_{j}$. Hence the Riemannian structure of $B$ must be the same as that inherited from $M$! Note that in our derivation, we did not impose this condition: it is a consequence of integrability.

In the general case of a non-vanishing $M$, we do not have the nice properties are are consequences of $B$ lying as a 1-parameter family of submanifolds in $M$. In particular, this means that when we use coordinates to do calculations (which we will not), it is wrong to claim that a certain $x^{i}$  furnish a coordinate on $B$, and with the additional coordinate $x^{0}$ this becomes a coordinate on $M$, and that the curvatures can be calculated in the usual way and they agree on $B$.

\subsection{Connection and curvature for the base space}
\label{torsion}

Let us now use $\omega^{i}$ as a frame on the base $B$. This is legitimate since we have a Riemannian submersion. We have
\begin{align*}
d\omega^{i}&=\omega^{j}\wedge\omega^{i}{}_{j}+\omega^{0}\wedge\omega^{i}{}_{0}\\
&=\omega^{j}\wedge\omega^{i}{}_{j}+\omega^{0}\wedge(K_{i}\omega^{0}+M_{ij}\omega^{j})\\
&=\omega^{j}\wedge(\omega^{i}{}_{j}-M^{i}{}_{j}\omega^{0}).
\end{align*}
The usual first set of Maurer-Cartan equations is, on the other hand,
\[
d\omega^{\mu}=\omega^{\nu}\wedge\omega^{\mu}{}_{\nu}.
\]
Comparing the two, we see that the first structural equation is \emph{not} satisfied. If on $B$, we continue to use the forms $\omega^{i}{}_{j}$, we have a Cartan connection with torsion, with torsion 2-form given by
\[
\widetilde\Omega^{i}=-M^{i}{}_{j}\omega^{j}\wedge\omega^{0}.
\]
This is the final interpretation of the quantity $M_{ij}$ alluded to earlier. Now, even though the forms $\omega^{i}{}_{j}$ fails to be the usual connection (in the sense of Levi-Civita) on the base manifold $B$, the forms $\omega^{i}$ are definitely legitimate moving coframes on $B$. Given the Riemannian structure of $M$ and an explicit form of a rigid flow, the Riemannian structure on $B$ is completely determined. Hence, there exists unique torsion free connection on $B$. In fact, due to uniqueness, such connection is easy to find. We just define the new connection as
\[
\widetilde\omega^{i}{}_{j}=\omega^{i}{}_{j}-M^{i}{}_{j}\omega^{0}
\]
and we have, now trivially
\[
d\omega^{i}=\omega^{j}\wedge\widetilde \omega^{i}{}_{j}.
\]
Note that this is the same thing as our ``absorption of inessential torsion'' using Cartan's equivalence method. Indeed, this is the geometrical meaning of the procedure of absorbing inessential torsion in the case of a Riemannian submersion.

We can now calculate the curvature of the base $B$:
\[
\widetilde\Omega^{i}{}_{j}=d\widetilde\omega^{i}{}_{j}+\widetilde\omega^{i}{}_{k}\wedge\widetilde\omega^{k}{}_{j}
\]
or
\[
\tfrac{1}{2}\widetilde R^{i}{}_{jkl}\omega^{k}\wedge\omega^{l}=d\widetilde\omega^{i}{}_{j}+\widetilde\omega^{i}{}_{k}\wedge\widetilde\omega^{k}{}_{j}.
\]

We want to express everything in terms of the intrinsic geometry of $L_{p}$ and $B$, together with the quantities $M_{ij}$ and $K_{i}$. By doing this calculation in the reduced bundle, the horizontal forms are $\omega^{0}$ and $\omega^{i}$, whilst $\widetilde\omega^{i}{}_{j}$ are vertical (and $\omega^{i}{}_{j}$, without the tilde, is a mixture of both).
The covariant derivative we will be using is \emph{not} the usual $\nabla$ defined by the set $\omega^{\mu}$ and $\omega^{\mu}{}_{\nu}$, but instead defined by the set $(\omega^{0},\omega^{i})$ and $(0,\widetilde\omega^{i}{}_{j})$. If we denote this derivative by $D$, then symbolically we have
\[
D=\nabla^{(L_{p})}\oplus\nabla^{(B)}
\]
where $\nabla^{L_{p}}$ and $\nabla^{B}$ are the Levi-Civita (torsion-free, metric compatible) with the induced metric  on $L_{p}$ and $B$, respectively. We will use semicolon ($;$) to denote components of covariant derivatives with respect to $D$, not $\nabla$. Let us first calculate derivatives of $K_{i}$ and $M_{ij}$, as these will be extensively used ($\dot A$ means $D_0A$):
\[
DK_{i}\equiv K_{i;j}\omega^{j}+\dot K_{i}\omega^{0}=dK_{i}-K_{j}\widetilde\omega^{j}{}_{i}=dK_{i}-K_{j}\omega^{j}{}_{i}+K_{j}M^{j}{}_{i}\omega^{0}
\]
and
\begin{align*}
DM_{ij}&\equiv M_{ij;k}\omega^{k}+\dot M_{ij}\omega^{0}
=dM_{ij}-M_{kj}\widetilde\omega^{k}{}_{i}-M_{ik}\widetilde\omega^{k}{}_{j}\\
&=dM_{ij}-M_{kj}\omega^{k}{}_{i}-M_{ik}\omega^{k}{}_{j}+M_{kj}M^{k}{}_{i}\omega^{0}+M_{ik}M^{k}{}_{j}\omega^{0}.
\end{align*}

We can obtain equations relating the intrinsic geometry and $M_{ij}$, $K_{i}$ by differentiating $\omega^{i}{}_{j}$. We have:

\begin{align*}
  \tfrac{1}{2}R^{i}{}_{j\mu\nu}\omega^{\mu}\wedge\omega^{\nu}&=d\omega^{i}{}_{j}+\omega^{i}{}_{k}\wedge\omega^{k}{}_{j}+\omega^{i}{}_{0}\wedge\omega^{0}{}_{j}\\
&=(-M^{i}{}_{j;k}+K^{i}M_{jk}-K_{j}M^{i}{}_{k}-K_{k}M^{i}{}_{j})\omega^{0}\wedge\omega^{k}\\
&\quad+(\tfrac{1}{2}\widetilde R^{i}{}_{jkl}-M^{i}{}_{k}M_{jl}+M^{i}{}_{j}M_{lk})\omega^{k}\wedge\omega^{l}.
\end{align*}
From which we learn, by comparing coefficients for the $\omega^{k}\wedge\omega^{l}$ term:
\begin{equation}
\label{gooaffa}    R_{ijkl}=\widetilde R_{ijkl}-M_{ik}M_{jl}+M_{il}M_{jk}+2M_{ij}M_{lk}.
\end{equation}
We could also obtain Bianchi identities by calculating in details $d^{2}\omega^{0}=0$:
\begin{align*}
  0&=d^{2}\omega^{0}\\
&=d(\omega^{i}\wedge\omega^{0}{}_{i})\\
&=(-K_{i;j}-\dot M_{ij}+K_{j}K_{i}+M_{ki}M^{k}{}_{j}+M_{kj}M^{k}{}_{i})\omega^{0}\wedge\omega^{i}\wedge\omega^{j}\\
&\quad+(-M_{ij;k}+K_{i}M_{jk})\omega^{i}\wedge\omega^{j}\wedge\omega^{k}.
\end{align*}
hence, by comparing coefficients of the $\omega^{0}\wedge\omega^{i}\wedge\omega^{j}$ term,
  \begin{equation}
    \label{eq:16}
0=-K_{i;j}-\dot M_{ij}+K_{j}K_{i}+M_{ki}M^{k}{}_{j}+M_{kj}M^{k}{}_{i}.
  \end{equation}
Antisymmetrise \eqref{eq:16} gives:
\begin{equation}
\label{confusion}
K_{[i;j]}=-\dot M_{ij}.
\end{equation}
In proposition \ref{rfif}, we see that a rigid motion is also isometric if and only if $K_{[i;j]}=0$. Hence, we have obtained the following, much easier to understand criteria for a rigid isometric motion:
\begin{thm}\label{finalc}
  A rigid motion is also isometric if and only if $\dot M_{ij}=0$ and $K_{i}$ remains constant along each leaf.
\end{thm}
In view of the interpretation of $M_{ij}$ as rotation coefficients, the above amounts to saying that $M_{ij}$ has to be constant along the flow if we have isometric motion.

\section{Generalisations to the Herglotz-Noether theorem}

Let us now come to homogeneous spacetime. Consider \eqref{gooaffa}. We have:
\begin{equation}
\label{pow1}\mathcal{R}_{ijkl}=-M_{ik}M_{jl}+M_{il}M_{jk}+2M_{ij}M_{lk}
\end{equation}
where $\mathcal{R}_{ijkl}=\widetilde R_{ijkl}-R_{ijkl}$ is a quantity that has the same symmetry as $\widetilde R_{ijkl}$. In particular, in the flat case, the two are equal. In homogeneous spacetimes, though they are not equal, $\mathcal{R}_{ijkl}$ is still defined on $B$, as $\widetilde R_{ijkl}$ is. This is not true for more general spacetimes, since $R_{ijkl}$ may then vary on each leaf. But then
\[
\mathcal{R}_{ijji}=2M_{ij}M_{ij}-M_{ij}M_{ji}+M_{ii}M_{jj}=3(M_{ij})^{2}\qquad\text{(no summation on $i,j$)}.
\]Since in the homogeneous case, $\mathcal{R}_{ijkl}$ is well-defined on $B$, we must have $\dot {\mathcal{R}}_{ijkl}=0$. This  immediately gives $\dot M_{ij}=0$.

We can now do the same thing as we did in the last section, but now aiming to calculate an expression for $R_{0ijk}$. We get
\[
R_{0ijk}=M_{i[j;k]}-2K_{i}M_{kj}
\]
Note that $R_{0ijk}$ vanishes by our assumption, and both $M_{i[j;k]}$ and $M_{kj}$ depends on only data given on the base. Hence if $M_{kj}$ has a non-zero component, to ensure this constraint it is necessary that $K_{i}$ remains constant on each leaf as well. Then, in view of theorem \ref{finalc}, we have proved that
\begin{thm}
In homogeneous spacetime, any rotational rigid motion is necessarily isometric.  
\end{thm}

This is our generalisation to the classical Herglotz-Noether theorem \cite{herglotz1910,noether1910,pirani1962,pirani1964,Giulini:2006p66}, which is restricted to the case of 4-dimensional rotational case. What about the converse? The converse is actually much easier to show, as has been done in \cite{pirani1962,pirani1964}. In our approach, the converse would seem actually trivial, which we will very briefly outline. Indeed, suppose $M$ is a homogeneous spacetime. Then the principal bundle on $M$ is itself a Lie group (e.g., in the case of Minkowski spacetime, the bundle itself is the Poincar\'e{} group). Let $\omega^{0}$ be a 1-form corresponding to an isometry on $M$. It lifts to the bundle to a form, which we will also denote by $\omega^{0}$. Since we are considering isometry, $\omega^{0}$ is an invariant Maurer-Cartan form on the bundle considered as a group. If we complete $\omega^{0}$ with $\omega^{i}$ and $\omega^{\mu}{}_{\nu}$, we have the canonical Maurer-Cartan coframe on the Lie group. This shows that in the structural equations
\[
d\omega^{A}=C^{A}_{BC}\omega^{B}\wedge\omega^{C}
\]
the functions $C^{A}_{BC}$ are constants, the structural constants of the Lie group, where now the indices $A, B, C$ takes value in single or double indices $0,i,j,k,\{ij\},\dots$. Doing the obvious manipulation then shows that $M_{ij}$ is antisymmetric (due to the antisymmetry of the structural constants) and \emph{constant}, which proves rigidity. Hence, we have the following
\begin{cor}
  In homogeneous spacetime, rotational rigid motion are in 1-1 correspondence with rotational isometric motions satisfying the relevant restrictions (e.g., being timelike, etc.). Hence the classification of rotational rigid motions in homogeneous spacetime is the same as the classification of rotational isometries, i.e., Killing vector fields.
\end{cor}

We must point out here that the theorem is only useful when applied to local flow. Indeed, suppose we have, for an inertial frame in a Minkowski spacetime, a global rotational flow, with no translation, boosts, etc., then in this frame, the velocity of particles measured in this frame is proportional to the distance from the centre of the rotation. Hence it is unbounded, and sooner or later it will exceed the speed of light $c$. Then our assumption that the flow is everywhere timelike becomes false. This argument is actually general and shows that no global rotating rigid motion can exist in Minkowski spacetime (the non-existence of the relativistic rotating rigid disk).
On the other hand, if we consider anti-de Sitter spacetime, unlike the previous case global rotation \emph{can} exist due to the (usual Riemannian) curvature of anti-de Sitter spacetime. However, an upper bound for the magnitude of $M_{ij}$ can easily be calculated such that the flow everywhere remains timelike. This might have interesting consequences in the AdS/CFT correspondence \cite{Bhattacharyya:2007p31}, where it gives bounds on the magnitude of black holes.

\section{Conclusion}

In this paper we have extended the classical Herglotz-Noether theorem to all dimensions. The relationship between Born-rigidity and isometry, as dictated by our theorem, is of valuable use to applications of fluid mechanics, and we mentioned the applications to the AdS/CFT correspondence. Further extensions of the theorem and theoretical background, as well as a general framework for dealing with all problems of structure-preserving submersion, is the subject of the paper \cite{me:001}.

\section*{Acknowledgement}
I would like to thank Prof Gary W Gibbons for valuable discussions concerning the problems of rigid motion and AdS/CFT correspondence.

\newpage
\nocite{me:001}
\nocite{Cartan:1983p6178}
\nocite{Cartan:2001p6159}
\nocite{pirani1962}
\nocite{Wahlquist:1967p2556}
\nocite{Wahlquist:1966p2548}
\nocite{Estabrook:1964p2608}
\nocite{Giulini:2006p66}
\nocite{Boyer:1965p2926}
\nocite{Robinson:1983p3530}
\nocite{tensor}
\nocite{Alvarez:2009p4142}
\nocite{Fels:1999p7361}
\nocite{Fels:1998p7335}
\nocite{JOlver:1995p9015,LBryant:1991p8950}

\bibliography{pub1}{}
\bibliographystyle{hplain}
\addcontentsline{toc}{section}{References}
\end{document}